\documentclass[preprintnumbers,superscriptaddress,nofootinbib]{revtex4}
\usepackage{amssymb}
\usepackage{amsmath}
\usepackage[dvips]{graphicx,psfrag}

\renewcommand{\thefootnote}{\#\arabic{footnote}}
\newcommand{\slashp}{\not{\hbox{\kern-3pt $P$}}}
\newcommand{\slashs}{\not{\hbox{\kern-3pt $S$}}}

\begin{document}

\renewcommand{\thefootnote}{\fnsymbol{footnote}}
\setcounter{footnote}{0}

\def\thefootnote{\fnsymbol{footnote}}

\preprint{CALT-68-2416}

\title{Grand Unification and Time Variation of the Gauge Couplings
\footnote{talk given by X. Calmet at the 10th International
Conference on Supersymmetry and Unification of Fundamental Interactions
(SUSY02), Hamburg, Germany, 17-23 June 2002.}}

\author{X. Calmet}
\email{calmet@theory.caltech.edu}
\affiliation{California Institute of Technology, Pasadena, 
California 91125, USA}
\author{H. Fritzsch}
\email{fritzsch@mppmu.mpg.de}
\affiliation{Sektion Physik 
Universit\"at M\"unchen, Theresienstr. 37A, D-80333 M\"unchen}
\begin{abstract}
Astrophysical indications that the fine structure constant is time
     dependent are discussed in the framework of grand unification
     models. A variation of the electromagnetic coupling constant
     could either be generated by a corresponding time variation of
     the unified coupling constant or by a time variation of the
     unification scale, or by both.  The case in which the time
     variation of the electromagnetic coupling constant is caused by a
     time variation of the unification scale is of special
     interest. It is supported in addition by recent hints towards a
     time change of the proton-electron mass ratio. Possible
     implications for baryogenesis are discussed.
\end{abstract}

\maketitle

The study of a possible time variation of the fundamental parameters
like for example the fine structure constant, has a long history that
can be traced back to Dirac \cite{dirac}. Some extensions of the
standard model of particle physics and in particular models that
couple gravity to the standard model, e.g. quintessence
\cite{Wetterich:2002ic}, require or at least allow a time dependence
of the parameters of the model. Our motivation to study the
implications of a time dependence of the fundamental parameters for
particle physics is not that much of a theoretical origin but rather
because there are indications
\cite{Webb:2000mn,Murphy:2002jx,Webb:2002vd,Murphy:2002ve,Potekhin:1998mf,Ivanchik:2002me}
coming from different astrophysical measurements that the parameters
of the standard model could be time dependent. We shall consider the
implications of Webb {\it et al.}'s result which indicate a possible
time dependence of the fine structure constant $\alpha$ and make
predictions that could be tested in other measurements.  If
interpreted in the simplest way, the data suggest that $\alpha$ was
lower in the past:
\begin{equation} \label{exinput}
\Delta \alpha / \alpha = (-0.72 \pm 0.18) \times 10^{-5}
\end{equation}
for a redshift $z \approx 0.5 \ldots 3.5$ \cite{Webb:2000mn}. We note
that since then the significance has increased $\Delta \alpha / \alpha
= (-0.57 \pm 0.10) \times 10^{-5}$
\cite{Murphy:2002jx,Webb:2002vd,Murphy:2002ve}. An interpretation as a
spatial dependence of fundamental parameters is also conceivable see
e.g. \cite{Rafelski:2002gq}.

It should be clear that Webb {\it et al.}'s result can
only be tested in a given theoretical framework. Therefore a negative
result in another sector cannot rule out that measurement, but can
basically only rule out a set of assumptions made about the model.

It would be surprising if only $\alpha$ was time dependent and we thus
expect that other parameters of the standard model should be time
dependent too. Unfortunately the standard model has too many
parameters which are uncorrelated. This implies a large number of time
dependent functions and makes the discussion of time dependence of the
fundamental constants in that framework not very efficient. We thus
consider grand unified models where one has relations between the
different parameters. We shall concentrate on SU(5) with ${\cal N}=1$
supersymmetry and S0(10) broken directly to the standard model where
unification is still possible due to thresholds effects
\cite{Lavoura:su}. The results are quite different which is of great
theoretical interest since a possible time variation of the parameters
could allow to test the ideas of grand unification.
 
We make the following assumptions:
\begin{itemize}
  \item The standard model is embedded into a grand unified theory.
  \item Unification takes place at all time.  
\item The physics
  responsible for the time evolution of the parameters only affects
  the unified coupling constant $\alpha_u$ and the unification scale
  $\Lambda_G$.  
\item The Yukawa couplings are time-independent at $\Lambda_G$. 
\item The Higgs bosons vacuum expectation values are
  time-independent at $\Lambda_G$.  
\end{itemize}
Our predictions only test the data coming from astrophysics together with this set of assumptions.

Assuming $\alpha_{u} = \alpha_u (t)$ and $\Lambda_G=\Lambda_G(t)$ and
 using the one loop renormalization group equations one finds:

\begin{eqnarray}
  \frac{1}{\alpha_i} \frac{\dot{\alpha}_i}{\alpha_i}=
  \left[\frac{1}{\alpha_u} \frac{\dot{\alpha}_u}{\alpha_u} 
 - \frac{b_i}{2 \pi} \frac{\dot{\Lambda}_G}{\Lambda_G}
  \right]
\end{eqnarray}
where $b^{{SM}}_i\!\!\!=(b^{{SM}}_1, b^{{SM}}_2, b^{{SM}}_3)=(41/10,
-19/6, -7)$ are the coefficients of the renormalization group
equations for the standard model and $b^{{S}}_i\!\!\!=(b^{{S}}_1,
b^{{S}}_2, b^{{S}}_3)=(33/5, 1, -3)$ are the coefficients of the
renormalization group equations in the ${\cal N}=1$ supersymmetric
case. This leads to
\begin{eqnarray} \label{eq3A}
  \frac{1}{\alpha} \frac{\dot{\alpha}}{\alpha}= \frac{8}{3}
\frac{1}{\alpha_s} \frac{\dot{\alpha}_s}{\alpha_s} -\frac{1}{2 \pi}
\left(b_2+\frac{5}{3} b_1-\frac{8}{3} b_3\right)
\frac{\dot{\Lambda}_G}{\Lambda_G}.  
\end{eqnarray}

We first consider the SU(5) supersymmetric case.  One may consider
different scenarios. We first keep $\Lambda_G$ invariant and consider
the case where $\alpha_u =\alpha_u (t)$ is time dependent. One gets
\cite{Calmet:2001nu}
\begin{eqnarray}
\frac{\dot{\Lambda}}{\Lambda}= -\frac{3}{8} \frac{2 \pi}{b_3^{SM}}
  \frac{1}{\alpha} \frac{\dot{\alpha}}{\alpha}= R
  \frac{\dot{\alpha}}{\alpha}, 
\end{eqnarray}
where $\Lambda$ is the QCD scale.  If we calculate
$\dot{\Lambda}/\Lambda$ using the relation above in the case of 6
quark flavors, neglecting the masses of the quarks, we find $R \approx
46$. There are large theoretical uncertainties in $R$. Taking thresholds
into account one gets $R=37.7\pm 2.3$ \cite{Calmet:2001nu}. The
uncertainty in $R$ is given, according to $\Lambda = 213^{+38}_{-35}
\mbox{MeV}$, by the uncertainty in the ratio $\alpha/\alpha_s$, which
is dominated by the uncertainty in $\alpha_s$.

We now consider the case where $\alpha_u$ is invariant, but $\Lambda_G
=\Lambda_G (t)$ is time dependent. One gets \cite{Calmet:2002ja}
\begin{eqnarray}
\frac{\dot{\Lambda}}{\Lambda}=
\frac{b_3^{S}}{b_3^{SM}} \left[ \frac{-2 \pi}{b_2^S+\frac{5}{3}
b_1^S}\right] \frac{1}{\alpha} \frac{\dot{\alpha}}{\alpha} \approx
-30.8 \frac{\dot{\alpha}}{\alpha}. 
\end{eqnarray}
It is interesting to notice that the effects of a time variation of
 the unified coupling constant or of a time variation of the grand
 unified scale are going in opposite directions.  Clearly those are
 two extreme cases and a time variation of both parameters is
 conceivable. Another possibility is dynamics between the grand
 unification scale and low energy physics \cite{grojean}. In that case it is
 conceivable to have a time variation of $\alpha$ but no time
 variation in the QCD sector. Such an effect could also be achieved in
 our approach by a fine tuning of the parameters $\alpha_u$ and
 $\Lambda_G$.

In a grand unified theory, the grand unified scale and the unified
  coupling constant may be related to each other via the Planck scale
  e.g. 
\begin{eqnarray} 
\frac{1}{\alpha_u}= \frac{1}{\alpha_{Pl}} + \frac{b_G}{2 \pi}
  \ln\left( \frac{\Lambda_{Pl}}{\Lambda_G} \right) 
\end{eqnarray}
  where $\Lambda_{Pl}$ is the Planck scale, $\alpha_{Pl}$ the value of
  the grand unified coupling constant at the Planck scale and $b_G$
  depends on the grand unified group under consideration.  This leads to
\begin{eqnarray}
  \frac{\dot{\alpha}}{\alpha} = \frac{- b_3^{SM}}{2 \pi} \left (
   \frac{ \frac{8}{3}b_G - b_2^S-\frac{5}{3} b_1^S}{ b_G-
   b_3^S}\right) \alpha \frac{\dot{\Lambda}}{\Lambda},  
\end{eqnarray}
in which case a test of the nature of the grand unified group is in
principle possible. It should be mentioned that the scale of
supersymmetry could also vary with time. One obtains:
\begin{eqnarray} \frac{1}{\alpha_i}
\frac{\dot{\alpha}_i}{\alpha_i}&=& \left[\frac{1}{\alpha_u}
\frac{\dot{\alpha}_u}{\alpha_u} - \frac{b_i^S}{2 \pi}
\frac{\dot{\Lambda}_G}{\Lambda_G} \right] + \frac{1}{2 \pi}
(b_i^S-b_i^{SM}) \frac{\dot{\Lambda}_S}{\Lambda_S}
\theta(\Lambda_S-\mu).  
\end{eqnarray} 
However without a specific model for supersymmetry breaking relating
the supersymmetry breaking scale to e.g. the grand unified scale, this
expression is not very useful since it only introduces a new unknown
function in the discussion.

The case in which the time variation of $\alpha$ is related to a time
variation of the unification scale is of particular
interest. $\Lambda_G$ could be related in specific models to vacuum
expectation values of scalar fields. Since the universe expands, one
might expect a decrease of the unification scale due to a dilution of
the scalar field. A lowering of $\Lambda_G$ implies according to
(\ref{eq3A}):
\begin{eqnarray} 
\frac{\dot{\alpha}}{\alpha}= - \frac{1}{2 \pi}
\alpha \left(b_2^S+\frac{5}{3} b_1^S\right)
\frac{\dot{\Lambda}_G}{\Lambda_G}= -0.014
\frac{\dot{\Lambda}_G}{\Lambda_G}.
\end{eqnarray}
If $\dot{\Lambda}_G / \Lambda_G$ is negative, $\dot{\alpha} / \alpha$
increases in time. That is consistent with the experimental
observation.  Taking $ \Delta{\alpha} / \alpha=-0.72 \times 10^{-5}$,
we would conclude $\Delta{\Lambda}_G / \Lambda_G=5.1 \times 10^{-4}$,
i.e. the scale of grand unification about 8 billion years ago was
about $8.3 \times 10^{12}$ GeV higher than today.

Monitoring the ratio $\mu=M_p/m_e$ could allow to see an effect.
Measuring the vibrational lines of H$_2$, a small effect was seen
recently: $\Delta \mu / \mu = (5.7 \pm 3.8) \times 10^{-5}$
\cite{Potekhin:1998mf}.  Supersymmetric SU(5) predicts $\frac{\Delta
\mu}{\mu} = 22 \times 10^{-5}$ with a rather large theoretical
uncertainty.  It is interesting that the data suggests that $\mu$ is
indeed decreasing, while $\alpha$ seems to increase. If confirmed,
this would be a strong indication that the time variation of $\alpha$
at low energies is caused by a time variation of the unification
scale. We would like to emphasize that our calculation is based on the
assumption that the proton mass is mainly determined by $\Lambda$. In
particular, we neglect the possible time changes of the electron mass
or of the quarks masses.

Under a further assumption, namely that $\dot \alpha/ \alpha$ is
constant, tests could be performed in quantum optics. We consider the
case in which $\Lambda(t)$ is time dependent. If the rate of change is
extrapolated linearly, $\Lambda_G$ is decreasing at a rate
$\frac{\dot{\Lambda}_G}{\Lambda_G}=-7\times 10^{-14}/$yr. The magnetic
moments of the proton $\mu_p$ as well as that of nuclei would increase
according to $\frac{\dot{\mu}_p}{\mu_p} = 30.8
\frac{\dot{\alpha}}{\alpha} \approx 3.1 \times 10^{-14}/ {\mbox yr}$.
The wavelength of the light emitted in hyperfine transitions, e.g. the
ones used in the cesium clocks being proportional to $\alpha^4
m_e/\Lambda$ will vary in time like $\frac{\dot{\lambda}_{hf}
}{\lambda_{hf}} = 4 \frac{\dot \alpha}{\alpha} -\frac{\dot
\Lambda}{\Lambda}\approx 3.5 \times 10^{-14}/\mbox{yr}\ \mbox{taking}\
\dot{\alpha}/\alpha\approx 1 \times 10^{-15}/\mbox{yr}$. The
wavelength of the light emitted in atomic transitions varies like
$\alpha^{-2}$: $\frac{\dot{\lambda}_{at} }{\lambda_{at}} = -2
\frac{\dot{\alpha} }{\alpha}$. One has ${\dot{\lambda}_{at}
}/{\lambda_{at}}\approx -2.0\times 10^{-15}/$yr. A comparison gives:

\begin{eqnarray}
  \frac{\dot{\lambda}_{hf}/\lambda_{hf}}{\dot{\lambda}_{at}/\lambda_{at}} = 
  -\frac{ 4 \dot{\alpha}/ \alpha - \dot \Lambda / \Lambda}{2 \dot{\alpha}/ \alpha }
  \approx -17.4.
\end{eqnarray}

It should be clear that our results are strongly model dependent. For
example in SO(10) without supersymmetry, varying the grand
unification scale, one finds:
\begin{eqnarray} 
\frac{\dot \Lambda }{\Lambda} = \left[ \frac{-2
\pi}{b_2^{SM} +\frac{5}{3} b_1^{SM}}\right] \frac{1}{\alpha}
\frac{\dot{\alpha}}{\alpha} =-234.8 \frac{\dot \alpha }{\alpha},
\end{eqnarray}
neglecting the threshold corrections. But, this model dependence is
what makes a possible time variation of the fundamental parameters so
interesting. In principle, we could test grand unified theories
without seeing any particle from a grand unified model. A time
variation of the gauge couplings could thus provide a new condition
for a grand unified theory besides reasonable proton decay and the
unification of the coupling constants. But, it would require a more
careful approach: calculations should take thresholds effects into
account and would become quite complicated. Furthermore it would
require measuring different time dependent quantities.

Clearly there are many constraints coming from different sectors and
different redshifts. See \cite{Uzan:2002vq} for a review. One
important constraint is the Oklo phenomenon which allows to derive a
severe constraint for the time variation of $\alpha$ during the last
two billion years ago \cite{Damour:1996zw}. But, this analysis is
performed under the assumption that only $\alpha$ is time
dependent. As we have shown the effects could be much larger in QCD
but go in the opposite direction. Some partial cancellation could
possibly take place. But, it has been shown that extracting a limit
for a time variation of the strong coupling constant is not an easy
task \cite{Beane:2002vq}.  Clearly it would be difficult to rule out
the results coming from astrophysics using data coming from a later
time. But, it would be very surprising if no effect was observed at a
previous time. Such effects could show up in the cosmic microwave
background \cite{Martins:2002qw}. Nucleosynthesis also allows to
constrain severely time variations of fundamental parameters (see
e.g. \cite{Nollett:2002da}).

Another question is what is really measured in
\cite{Webb:2000mn,Murphy:2002jx,Webb:2002vd,Murphy:2002ve}. The
authors use a so-called many multiplet method, and fit the
relativistic correction $\Delta=(Z\alpha)^2 \left ( \frac{1}{j+1/2} -C
\right)$ taking $C \sim 0.6$ calculated from QED using a many-body
method \cite{Murphy:2002jx}.  But, the question is at what level would
QCD affect this measurement? Most probably, the splitting of the
observed spectral lines is only due to the QED atomic structure. Only
inner shells of heavy atoms are sensitivity to the nuclear size
influenced by the QCD parameter. But, a change in nuclear size, and
thus of the charge distribution would at some level impact the
magnitude of the parameter $C$. So it is a justified question to ask
what is exactly measured if the effect is really much stronger in QCD,
as expected from grand unified models.

We shall finally discuss the possible implications for baryogenesis of
a time dependence of fundamental parameters. The aim of baryogenesis
is to explain the matter/anti-matter asymmetry. Any model must fulfill
the Sakharov's conditions:
\begin{itemize} 
\item[a)] The baryon number must be violated by some process (the net
baryon number must change over time).
\item[b)] $C$ and $CP$ must be violated (no perfect equality between rates
of $\Delta B \neq 0$ processes otherwise no asymmetry could evolve
from initially symmetric state).
\item[c)] A departure from thermal equilibrium is required, otherwise
$CPT$ would assure compensation between processes increasing or
decreasing the baryon number.
\end{itemize}

The standard model has a problem with points $b)$ and $c)$. It has not
enough $CP$ violation and the Higgs boson mass is too high to have a
first order phase transition. The Higgs boson mass $m_H$ should be
smaller than 40 GeV (see e.g. \cite{Carena:1997ki}). This is actually
a constraint on $\lambda_H$, the Higgs self-coupling. The condition
for point $c)$ is $E_B/\lambda(T_C) \ge 1$ which is equivalent to (see
e.g. \cite{Carena:1997ki}):
\begin{eqnarray}
\sqrt{2} \frac{2
M_W^3 +M_Z^3}{3 \pi v^3 \lambda(T_C) } \ge 1.
\end{eqnarray}

On the other hand the baryon number is proportional to (see
e.g.\cite{Farrar:sp}) $J=\sin(\theta_{12}) \sin(\theta_{13})
\sin(\theta_{23}) \sin(\delta_{CP})$ 
\begin{eqnarray}
n_B &\propto& (m_t^2 -m_c^2) (m_t^2 -m_u^2) (m_c^2 -m_u^2) \\ \nonumber && 
(m_b^2 -m_s^2) (m_b^2 -m_d^2) (m_s^2 -m_d^2) J/T^{12}\sim 10^{-21} . 
\end{eqnarray}
which is much smaller that the measured baryon number $\sim (4-10)
\times 10^{-11}$ (see e.g. \cite{Farrar:sp} for a discussion).  Thus
the standard model is not able to provide the right baryon number and
a phase transition. But, things might slightly change if the
fundamental parameters of the standard model are time dependent. It
seems impossible to explain the baryon number in that framework in
view of the large discrepancy. But, if the parameters of the Higgs
potential were time dependent there could be a chance to get a phase
transition of 1st order. In general a parameter $p$ can be decomposed
into one $p=p_{gut}+p_{loop}(t)$, where $p_{gut}$ is the value of the
parameter at the grand unified scale and $p_{loop}(t)$ is the time
dependence induced by loop corrections. In our case we assumed that
the functions $p_{gut}$ for the Higgs and Yukawa sectors are time
independent. But, these parameters get a time dependence through the
radiative corrections which involve the coupling constants. An
estimate in SO(10) yields $\frac{\Delta v}{v}=0.4 \%$, $\frac{\Delta
\lambda_H}{\lambda_H}=-2 \%$ and $\frac{\Delta m_t}{m_t}=20 \%$ in
$10^{10}$ years, doing a linear extrapolation of the results of
\cite{Webb:2000mn}. The effect is stronger for the top mass than for
the $SU(2)$ Higgs sector because the top mass has a wild
running. Obviously the time variation of the parameters of the Higgs
potential obtained from that effect alone cannot explain the phase
transition. But, the time variation of the parameter $\lambda_H$ is
rather unconstrained by experiment. We could relax the assumption that
we made concerning the time invariance of $\lambda_H$ at the grand
unified scale thereby having a phase transition of 1st order in the
early universe.  Keeping $v$ roughly constant (notice that a time
variation of $v$ is strongly constrained by nucleosynthesis which is
very sensitive to $G_F$), the Higgs boson mass of the $SU(2)$ sector
could have been around $40$ GeV in the early universe if
$(\lambda_H)_{gut}$ is strongly time dependent. Clearly the physics of
the early universe could be affected by a time variation of the
fundamental parameters. It remains to see if any predictions can be
made in that framework. This will be difficult in view of the
potentially large number of uncorrelated time dependent functions.

{\sl Acknowledgment:} 
We shall like to thank J. Rafelski for enlightening discussions.

\end{document}